\documentclass[12pt]{article}

\usepackage{array}
\usepackage{amsfonts}

\def\Tr{\mathop{\rm Tr}}

\begin{document}

\title{Quite a Character:\\
The Spectrum of Yang-Mills on $S^3$\footnote{Based on the Brown
University Undergraduate Thesis {\it Lie Algebras and ${\mathcal{N}} = 4$
Yang-Mills Theory} by Taylor H.~Newton~'08.}}

\author{Taylor H. Newton and Marcus Spradlin\\
Department of Physics\\
Brown University\\
Box 1843\\
Providence, RI 02912 USA}

\date{October 2008}

\maketitle

\begin{abstract}
We introduce a simple method to extract the representation content
of the spectrum of a system with SU(2) symmetry from its partition function.
The method is easily generalized to systems with SO(2,4) symmetry,
such as conformal field theories in four dimensions.
As a specific application we obtain an explicit generating function
for the representation content of free planar Yang-Mills theory 
on $S^3$.  The extension to ${\cal N} = 4$ super Yang-Mills is also discussed.
\end{abstract}

\newpage

\section{Introduction and Summary}

Although the vast majority of
work on gauge/string duality has been conducted within the solid
framework of the AdS/CFT correspondence, the decades-old arguments
of 't Hooft and Polyakov suggest that general large $N$ gauge theories
should admit dual reformulations as string theories.
Indeed even the simplest large $N$ gauge theory, pure SU($N$) Yang-Mills,
exhibits a string-like Hagedorn density of states when the theory
is formulated on a sphere whose radius is small compared to the dynamically
generated scale,
where the partition
function may be reliably computed in perturbation
theory~\cite{Aharony:2003sx,Aharony:2005bq}
(Hagedorn behaviour in the supersymmetric case was first
discussed in~\cite{Sundborg:1999ue,HaggiMani:2000ru,Sundborg:2000wp}).

The present work was motivated by~\cite{Barabanschikov:2005ri},
where the representation content of free large $N$ Yang-Mills theory
on $S^3$, or equivalently the spectrum of particles in this theory's
as yet unknown
string dual, was determined.
Specifically, an explicit algorithm was presented for calculating the
number $N_{[d,j_1,j_2]}$ of times that SO(2,4) representation $[d,j_1,j_2]$
appears in the spectrum of the theory.  Here $d$ is an integer labelling
the energy of a state (in units of $1/R$) and $j_1,j_2$ are the two
spin quantum numbers arising from the SO(4) isometry group of $S^3$.
Although explicit, the result of~\cite{Barabanschikov:2005ri} is rather
complicated.  We present in eq.~(\ref{eq:mainpure}) a
relatively simple explicit formula for
the generating function
\begin{equation}
\label{eq:gdef}
G(q,a,b) = \sum_{d=1}^\infty \sum_{j_1,j_2 \in {\mathbb{N}}/2}
N_{[d,j_1,j_2]} q^d a^{j_1} b^{j_2}
\end{equation}
of the spectrum degeneracies.  As a check on our result,
we display the expansion of our generating function through
order $q^9$ in Table~1 and find perfect corroboration with the
corresponding results displayed in~\cite{Barabanschikov:2005ri}.

The key step in our construction follows from a straightforward
manipulation of SU(2) characters detailed in section 2 and extended
to the four-dimensional conformal group SO(2,4) in section 3.
In the final subsection of the paper we comment on the
possibility of extending
our approach to maximally supersymmetric ${\mathcal{N}} = 4$ Yang-Mills
theory.
Partition functions for this theory have been studied extensively, with
the most closely related work
including~\cite{Bianchi:2003wx,Beisert:2003te,Beisert:2004di,Bianchi:2004ww,Bianchi:2004xi}.
Let us note however that in contrast to those papers, where partition
functions counted numbers of (super)conformal primary {\it states},
here (as in~\cite{Barabanschikov:2005ri}) we focus on
the problem of counting
the number of (super)conformal primary {\it representations} in the spectrum.
To elucidate this distinction we point out that to count the number
of primary states in the SO(2,4) case would require including an additional
factor $(2j_1+1)(2j_2+1)$ inside the summand of eq.~(\ref{eq:gdef}).

\section{A Simple Relation for SU(2) Characters}

\subsection{The Relation}

We begin by considering a partition function
\begin{equation}
Z(a) = \Tr[a^{J_3}]
\end{equation}
for some system with a discrete spectrum exhibiting an
SU(2) symmetry.
Additional factors such as $x^H$ (where $H$ is any operator that
commutes with $J_3$) may be 
included inside the trace without
affecting the following arguments.
A consequence of SU(2) symmetry is that the function $Z(a)$ must be
expressible as a linear combination
\begin{equation}
\label{eq:z}
Z(a) = \sum_{j \in {\mathbb{N}}/2} N_j \chi_{[j]}(a),
\end{equation}
of non-negative integer coefficients $N_j$
and the characters
$\chi_{[j]}$
of the irreducible representations 
of SU(2).
We first restrict our attention to cases where this sum has only
a finite number of nonzero terms.
It is then clear from the explicit formula
\begin{equation}
\label{eq:chi}
\chi_{[j]}(a) = \sum_{m=-j}^{+j} a^j
= \frac{a^{j+1/2} - a^{-j-1/2}}{a^{1/2} - a^{-1/2}}
= \frac{\sin((j + 1/2) \theta)}{\sin (\theta/2)}, \qquad a = e^{i \theta}
\end{equation}
that $Z(a)$ admits a Laurent expansion in $a^{1/2}$
with only a finite number of nonzero coefficients.
Since each $\chi_{[j]}(a)$ is symmetric under $a \to 1/a$,
the partition function $Z(a)$ must have this symmetry as well, so
the Laurent expansion must take the form
\begin{equation}
\label{eq:laurent}
Z(a) = Z_0 + \sum_{k \in {\mathbb{Z}}^+/2} Z_k (a^k + a^{-k}) =
Z_0 + 2 \sum_{k \in {\mathbb{Z}}^+/2} Z_k  \cos(k \theta).
\end{equation}

Our goal is to find a simple way to find all of the coefficients $N_j$
given some generic partition function $Z(a)$ of this form.
Of course any individual
coefficient $N_j$ may be obtained by exploiting orthogonality
of the SU(2) characters,
\begin{equation}
\int_0^{4 \pi} \frac{\sin^2(\theta/2)}{2 \pi} d\theta
\ 
\overline{\chi_{[j]}(a)}
\chi_{[k]}(a)
 = \delta_{j,k},
\end{equation}
from which it follows that
\begin{equation}
\label{eq:nfromorth}
N_j = \int_0^{4 \pi} \frac{\sin^2(\theta/2)}{2 \pi} d\theta\ 
\overline{\chi_{[j]}(a)} Z(a).
\end{equation}

In order to encapsulate all of the $N_j$ simultaneously we find it
convenient to define
a generating function $G(a)$ for the $N_j$,
\begin{equation}
\label{eq:Gdef}
G(a) = \sum_{j \in {\mathbb{N}}/2} N_j a^j.
\end{equation}
Substituting eq.~(\ref{eq:nfromorth}) into eq.~(\ref{eq:Gdef})
and using eqs.~(\ref{eq:chi}),~(\ref{eq:laurent})
we find
\begin{equation}
\label{eq:Gresult}
G(a) = \sum_{j  \in {\mathbb{N}}/2} a^j \int_0^{4 \pi}
\frac{d \theta}{2 \pi}
\sin((j + 1/2) \theta)\sin(\theta/2) 
\Big[
Z_0 + 2 \sum_{k \in {\mathbb{Z}}^+/2} Z_k \cos (k \theta) \Big].
\end{equation}
Since the sum over $k$ involves (by assumption) only a finite number of
non-zero terms we are free to perform the integral first, using the result
\begin{eqnarray}
\int_0^{4 \pi}
\frac{d \theta}{2 \pi}
\sin((j + 1/2) \theta)\sin(\theta/2)
\cos (k \theta) &=&
\frac{1}{2} \delta_{j,k} - \frac{1}{2} \delta_{j+1,k}, \quad k > 0,\cr
&=& \delta_{j,0}, \qquad\qquad\qquad k =0.
\end{eqnarray}
Thus we arrive at
\begin{equation}
G(a) = \sum_{j \in {\mathbb{N}}/2} a^j \Big[ Z_0 \delta_{j,0}
+ \sum_{k \in {\mathbb{Z}}^+/2} (\delta_{j,k} - \delta_{j+1,k}) Z_k\Big].
\end{equation}
This equation, which is the central result of this section, may be
succinctly summarized as
\begin{equation}
\label{eq:central}
G(a) = \left\{ (1 - 1/a) Z(a) \right\}_{a^{\ge 0}}
\end{equation}
where the notation $a^{\ge 0}$ indicates that one should calculate
the Laurent expansion of the quantity in curly braces and then truncate
that expansion by keeping only those terms with non-negative powers of $a$.

We have established eq.~(\ref{eq:central}) by a rather complicated argument,
but it is straightforward to check that it is correct by directly
substituting eqs.~(\ref{eq:z}) and~(\ref{eq:chi}), which gives
\begin{eqnarray}
\label{eq:simple}
\left\{ (1 - 1/a) Z(a) \right\}_{a^{\ge 0}} &=& \Big\{
(1-1/a)
\sum_{j \in {\mathbb{N}}/2} N_j
\frac{a^{j+1/2} - a^{-j-1/2}}{a^{1/2} - a^{-1/2}}
\Big\}_{a^{\ge 0}}\cr
&=& \Big\{ \sum_{j \in {\mathbb{N}}/2} N_j
(a^j - a^{-j-1})
\Big\}_{a^{\ge 0}}
\cr
&=& \sum_{j \in {\mathbb{N}}/2} N_j a^j
\cr
&=& G(a).
\end{eqnarray}
Although we assumed above that $Z(a)$ had a finite Laurent expansion,
to ease our way through the proof, the simple argument presented
in eq.~(\ref{eq:simple}) demonstrates that eq.~(\ref{eq:central}) holds more
generally for formal power series.

\subsection{Restricting to Non-Negative Powers}

In this section we suggest a simple method to
explicitly implement the step of restricting to non-negative powers
of a Laurent expansion, which plays a crucial role in our
result~(\ref{eq:central}).

We first consider a function $f(z)$ with Laurent expansion
\begin{equation}
f(z) = \sum_{n=-\infty}^\infty c_n z^n.
\end{equation}
The basic identity we need involves the contour integral
\begin{equation}
\frac{1}{2 \pi i} \oint_{\cal C} dz \frac{z^n}{z-a},
\end{equation}
where $n$ is an integer and
${\cal C}$ is a contour around the origin with radius greater than $|a|$.
For $n \ge 0$ the result is $a^n$, by Cauchy's integral
formula, while for $n<0$ the result of integration is zero.
Therefore, in terms of the notation
introduced in eq.~(\ref{eq:central}) we find
\begin{equation}
\{ f(a) \}_{a \ge 0} =
\frac{1}{2 \pi i} \oint_{\cal C} dz \frac{f(z)}{z-a}.
\end{equation}

This analysis is valid for a function $f(z)$ whose Laurent
expansion has only integer powers of $z$.  In order to apply this
to an SU(2) partition function $Z(a)$ we need to modify it to allow
half-integer powers as well.  This is easily accomplished by taking
$a \to a^2$ before performing the contour integral, and then $a \to
\sqrt{a}$ to restore $a$ afterwards.  Also including the factor
$(1-1/a)$ which appears in eq.~(\ref{eq:central}), we find that our
result~(\ref{eq:central}) may be recast as the contour integral
\begin{equation}
G(a) =
\{ (1-1/a) Z(a) \}_{a^{\ge 0}}
= \frac{1}{2 \pi i} \oint_{\cal C} dz \frac{1 - 1/z^2}{z - \sqrt{a}}
Z(z^2),
\end{equation}
where ${\cal C}$ is a contour around the origin with radius greater
than $|a|^{1/2}$.

\section{Extension to SO(2,4)}

In this section our goal will be to extend the result of section 2 to the
conformal group in four dimensions, SO(2,4).
For a conformally invariant theory we can consider the partition
function
\begin{equation}
Z(q,a,b) = \Tr[q^H a^{J_3} b^{J_3'}]
\end{equation}
where $(H,J_3,J_3')$ are simultaneously commuting generators
of the maximal compact subgroup SO(2)$\times$SU(2)$\times$SU(2) of
SO(2,4).  In the application to Yang-Mills theory on
$S^3$ discussed below, $H$ will be the Hamiltonian
and the $J_3$'s will be generators
of the isometry group of $S^3$.

Conformal symmetry guarantees that the partition function may be
expressed as a linear combination
\begin{equation}
\label{eq:zconf}
Z(q,a,b) = \sum_{d,j_1,j_2} N_{[d,j_1,j_2]} \chi_{[d,j_1,j_2]}(q,a,b)
\end{equation}
of the characters of irreducible representations of SO(2,4).
The non-negative integer coefficient $N_{[d,j_1,j_2]}$ counts
the number of times the representation $[d,j_1,j_2]$ appears in the
the spectrum.
Our goal is to find a simple way to obtain, from a
given $Z(q,a,b)$, the generating function~(\ref{eq:gdef})
for these multiplicities $N_{[d,j_1,j_2]}$.

The analysis is somewhat complicated by the fact that
the irreducible representations $[d,j_1,j_2]$ of SO(2,4)
come in a couple of different varities (see for example~\cite{Mack:1975je}).
The generic (``long'') multiplet
has $d > j_1 + j_2 + 2$ and character
\begin{equation}
\label{eq:chiconf}
\chi_{[d,j_1,j_2]}(q,a,b) =
\frac{q^d \chi_{[j_1]}(a) \chi_{[j_2]}(b)}{
(1 - q x_1)(1 - q x_2) (1-q x_3)(1 - q x_4)},
\end{equation}
where
\begin{equation}
\label{eq:xi}
x_1 = \sqrt{a b}, \qquad
x_2 = \sqrt{a/b}, \qquad
x_3 = \sqrt{b/a}, \qquad
x_4 = 1/\sqrt{a b}.
\end{equation}
Let us first consider for simplicity the case
of a theory which has only long multiplets in its spectrum.
Then, we can substitute eq.~(\ref{eq:chiconf}) into eq.~(\ref{eq:zconf}),
multiply both sides by $\prod (1 - q x_i)$, and
apply the result~(\ref{eq:central}) separately for each SU(2)
to arrive at the following simple result
\begin{equation}
\label{eq:centralconf}
G(q,a,b) = \Big\{ (1 - 1/a) (1-1/b)
Z(q,a,b) \prod_{i=1}^4 (1 - q x_i) \Big\}_{a^{\ge 0}, b^{\ge 0}}.
\end{equation}
Here, the subscript $a^{\ge 0}, b^{\ge 0}$ is an instruction to
perform a Laurent expansion of the quantity in curly braces
in the variables $a$ and $b$, and then to
discard any terms containing a negative power of $a$ or $b$.

A generic conformal theory also has ``short'' multiplets
in its spectrum.  These come in two varieties,
one with character
\begin{equation}
\label{eq:shortone}
\overline{\chi}_{[d,j_1,j_2]} =
\chi_{[d,j_1,j_2]} - \chi_{[d+1,j_1-1/2,j_2-1/2]}
\end{equation}
for $d = j_1 + j_2 + 2$ and $j_1 j_2>0$, and the second with
character
\begin{equation}
\label{eq:shorttwo}
\overline{\chi}_{[d,j_1,j_2]} =
\chi_{[d,j_1,0]} - \chi_{[d+1,j_1-1/2,1/2]} +
\chi_{[d+2,j_1-1,0]}
\end{equation}
for $d = j_1 + 1$ and $j_1 \ge 0$ (or the same with the two SU(2)'s
interchanged).
If one were
to naively apply eq.~(\ref{eq:centralconf}) to a partition function $Z$ 
containing short multiplets, then some of the integer coefficients in
the resulting generating function $G$ would come out too small.
In the applications considered below we will first identify all
terms arising from short multiplets and then explicitly add back
the necessary terms to compensate for the missing contributions to the
generating function.
An entirely analagous operation has been applied to very closely
related manipulations on partition functions
in~\cite{Bianchi:2003wx,Beisert:2003te,Beisert:2004di,Bianchi:2004ww,Bianchi:2004xi}.

As an example of how this works in practice, let us
consider a theory with many long multiplets and a single short
multiplet of type~(\ref{eq:shorttwo}) with $j_1 = 1$.
In this case the desired generating function
$G(q,a,b)$ may be calculated as
\begin{equation}
\label{eq:stepone}
\Big\{ (1 - 1/a) (1-1/b)
\left[Z(q,a,b)
- \overline{\chi}_{[2,1,0]}(q,a,b)
\right] \prod_{i=1}^4 (1 - q x_i) \Big\}_{a^{\ge 0}, b^{\ge 0}}
+ q^2 a.
\end{equation}
That is, we first subtract the ``offending'' short representation $[2,1,0]$
from the partition function $Z$, leaving a partition function with
only long multiplets, enabling eq.~(\ref{eq:centralconf}) to be applied.
Finally we add back to $G(q,a,b)$ the term $+q^2 a$ which counts the
multiplet $[2,1,0]$.
A simple calculation then reveals that~(\ref{eq:stepone}) is equivalent to
\begin{equation}
\label{eq:conc}
\Big\{ (1 - 1/a) (1-1/b)
Z(q,a,b)
\prod_{i=1}^4 (1 - q x_i) \Big\}_{a^{\ge 0}, b^{\ge 0}}
+ (\sqrt{a b} - q) q^3.
\end{equation}
The conclusion of this analysis is that the result~(\ref{eq:centralconf})
may still be used to calculate the generating function $G$ for a partition
function $Z$ containing the short multiplet $[2,1,0]$ as long as the
compensating factor $+ (\sqrt{a b} - q) q^3$ is added afterward.
A similar analysis may be performed to find the
necessary
``compensating factor''
for all other short multiplets.

\section{Applications to Yang-Mills Theory on $S^3$}

The method described in section 3 can be applied
to any partition function $Z(q,a,b)$
with SO(2,4) symmetry.  We will now focus our attention on
two particular examples of such theories.  The first is pure
Yang-Mills theory on $S^3$ and the
second is the maximally supersymmetric ${\mathcal{N}} = 4$ Yang-Mills theory
on $S^3$.
Both theories will be considered in the planar limit, i.e.~with gauge
group U($N$) in the limit of infinite $N$, and with coupling constant
$\lambda = 0$.

\subsection{Pure Yang-Mills Theory}

As mentioned in the introduction, this theory has a dimensionless
coupling constant $\lambda = \Lambda R$ where $R$ is the radius of
the $S^3$ and $\Lambda$ is the dynamically generated scale.
When $\lambda = 0$,
the partition function of the theory may be calculated
exactly~\cite{Aharony:2003sx},
\begin{equation}
\label{eq:zpure}
Z(q,a,b) = \Tr[q^{H R} a^{J_3} b^{J_3'}] =
- \sum_{n=1}^\infty \frac{\phi(n)}{n}
\log\left[1 - z(q^n,a^n,b^n)\right].
\end{equation}
Here $H$ is the Hamiltonian of the theory on $S^3$,
which at $\lambda = 0$ may be identified with the dilatation operator
on ${\mathbb{R}}^4$, $J_3$ and $J_3'$ are generators of the two
SU(2) symmetry groups, and $\phi(n)$ is the Euler totient function which
counts the number of positive integers less than or equal to
$n$ which are coprime to $n$.  Finally, $z(q,a,b)$ is given by
\begin{equation}
z(q,a,b) = 1 + \frac{(q - q^3) (x_1 + x_2 + x_3 + x_4) + q^4 - 1}{
(1-q x_1)(1-q x_2)(1-q x_3)(1 - q x_4)}
\end{equation}
in terms of the $x_i$ defined in eq.~(\ref{eq:xi}).
Before proceeding we remark that the partition function for the theory
with gauge group SU($N$) rather than U($N$) may be obtained trivially
by adding $- z(q,a,b)$ to eq.~(\ref{eq:zpure}).

After a relatively simple analysis of the possible short multiplets,
we find that the desired generating function for pure Yang-Mills theory is
\begin{equation}
\label{eq:mainpure}
G(q,a,b) = \Big\{ (1 - 1/a) (1-1/b)
Z(q,a,b)
\prod_{i=1}^4 (1 - q x_i) \Big\}_{a^{\ge 0}, b^{\ge 0}}
+ C(q,a,b)
\end{equation}
where $Z$ is given in eq.~(\ref{eq:zpure}) and the compensating factor
\begin{equation}
C(q,a,b) = 
\frac{\sqrt{a b} (1 + \sqrt{a b} q + (a^2 + b^2) q^2) q^5}{1 - a b q^2}
+ 2 (\sqrt{a b} - q) q^3
\end{equation}
corrects for the short multiplets as explained in the previous section.
Here the first
term accounts for the infinite tower of short multiplets of
type~(\ref{eq:shortone})
while the second term accounts for the two short multiplets ($[2,1,0]$ and
$[2,0,1]$) of type~(\ref{eq:shorttwo}).

Formula~(\ref{eq:mainpure})
is the central result of this paper.
The generating function
$G(q,a,b)$ encapulates the number of times the
SO(2,4) representation $[d,j_1,j_2]$
appears in the spectrum of free Yang-Mills theory on $S^3$.  We can
read off the individual $N_{[d,j_1,j_2]}$ for some of the lowest lying states
by performing a series expansion in $q$.  The coefficient of $q^d$ is,
for each integer $d$, a polynomial $p_d$ in $a^{1/2}$ and $b^{1/2}$.
Through energy level $d=10$ we have calculated the representation content,
as shown in Table~1.
Our results are in perfect agreement with those
of~\cite{Barabanschikov:2005ri}.
In contrast to the explicit, but rather complicated formula presented
there, we find that the $N_{[d,j_1,j_2]}$ may be rather simply extracted
from eq.~(\ref{eq:mainpure}).

\begin{table}[ht]
\begin{center}
\begin{tabular}{|c|m{4.5in}|}
\hline
$d$&
$\qquad\qquad\qquad\qquad\qquad\qquad\quad p_d(a,b)$\tabularnewline
\hline
\hline
$1$&
$0$\tabularnewline
\hline
$2$&
$a+b$\tabularnewline
\hline
$3$&
$0$\tabularnewline
\hline
$4$&
$2+a^{2}+ab+b^{2}$\tabularnewline
\hline
$5$&
$a^{3/2}b^{3/2}$\tabularnewline
\hline
$6$&
$2+2a+a^{3}+2b+2ab+a^{2}b+a^{3}b+ab^{2}+a^{2}b^{2}+b^{3}+ab^{3}$\tabularnewline
\hline
$7$&
$4a^{3/2}b^{1/2}+2a^{5/2}b^{1/2}+4a^{1/2}b^{3/2}+4a^{3/2}b^{3/2}+2a^{5/2}b^{3/2}+2a^{1/2}b^{5/2}+2a^{3/2}b^{5/2}+a^{5/2}b^{5/2}$\tabularnewline
\hline
$8$&
$6+4a+5a^{2}+a^{3}+2a^{4}+4b+10ab+7a^{2}b+5a^{3}b+a^{4}b+5b^{2}+7ab^{2}+8a^{2}b^{2}+3a^{3}b^{2}+a^{4}b^{2}+b^{3}+5ab^{3}+3a^{2}b^{3}+a^{3}b^{3}+2b^{4}+ab^{4}+a^{2}b^{4}$\tabularnewline
\hline
$9$&
$14a^{1/2}b^{1/2}+20a^{3/2}b^{1/2}+15a^{5/2}b^{1/2}+6a^{7/2}b^{1/2}+20a^{1/2}b^{3/2}+28a^{3/2}b^{3/2}+18a^{5/2}b^{3/2}+7a^{7/2}b^{3/2}+2a^{9/2}b^{3/2}+15a^{1/2}b^{5/2}+18a^{5/2}b^{3/2}+7a^{7/2}b^{3/2}+2a^{9/2}b^{3/2}+15a^{1/2}b^{5/2}+18a^{3/2}b^{5/2}+12a^{5/2}b^{5/2}+4a^{7/2}b^{5/2}+6a^{1/2}b^{7/2}+7a^{3/2}b^{7/2}+4a^{5/2}b^{7/2}+a^{7/2}b^{7/2}+2a^{3/2}b^{9/2}$\tabularnewline
\hline
\end{tabular}
\end{center}
\caption{The spectrum of pure Yang-Mills theory on $S^3$
at zero coupling for energy levels $d=1$ through $d=10$ (in units of
the inverse radius of $S^3$).
The coefficient of $a^{j_1} b^{j_2}$ in the polynomial
$p_d(a,b)$ is the number of times that the SO(2,4) irreducible representation
$[d,j_1,j_2]$ appears in the spectrum.  These numbers were obtained
in~\cite{Barabanschikov:2005ri} by completely dissimilar means.}
\end{table}

\subsection{${\mathcal N} = 4$ Supersymmetric Yang-Mills Theory}

Maximally supersymmetric Yang-Mills theory is exactly conformal
for any value of the coupling constant.
The partition function of the free ($\lambda=0)$
supersymmetric theory on $S^3$ 
is given by the same formula~(\ref{eq:zpure}) as above, but with a
modified formula for $z$ which now takes the form
\begin{equation}
z(q,a,b)
= 1 + \frac{(q-q^3)(6+x_1+x_2+x_3+x_4) + q^4 - 1 + 4 q^{3/2} (1-q) y}
{(1-q x_1)(1-q x_2)(1-q x_3)(1-q x_4)},
\end{equation}
where the $x_i$ are still given by eq.~(\ref{eq:xi}) and
\begin{equation}
y = \sqrt{a} + \frac{1}{\sqrt{a}} + \sqrt{b} + \frac{1}{\sqrt{b}}.
\end{equation}
The first ${\cal O}(\lambda)$ correction to the free partition function
was calculated in~\cite{Spradlin:2004pp}.

Although we could
apply eq.~(\ref{eq:centralconf}),
with an appropriate compensating term to account for short
multiplets, the result would be only
partially satisfying.
The reason is that
the SO(2,4) conformal symmetry is only part of the much larger
PSU(2,2$|$4) symmetry group of the supersymmetric
theory.
A single irreducible representation of PSU(2,2$|$4) contains numerous
irreducible SO(2,4) representations (see~\cite{Dobrev:1985qv} as well as
the encyclopedic reference~\cite{Dolan:2002zh}),
each of which would be
counted separately
if we were to apply eq.~(\ref{eq:centralconf}).
In order to exploit the full symmetry we would like to consider a finer
partition function
\begin{equation}
Z(q,a,b,y_1,y_2,y_3) = \Tr[q^{H R}
a^{J_3} b^{J_3'}
y_1^{R_1} y_2^{R_2} y_3^{R_3}],
\end{equation}
which includes three chemical potentials for the SO(6) R-symmetry,
and then find the corresponding generating function
\begin{equation}
G(q,a,b,y_1,y_2,y_3) = \sum N_{[d,j_1,j_2,s_1,s_2,s_3]}
q^d a^{j_1} b^{j_2} y_1^{s_1} y_2^{s_2} y_3^{s_3}
\end{equation}
where $N_{[d,j_1,j_2,s_1,s_2,s_3]}$ counts the number of irreducible
representations with quantum numbers $[d,j_1,j_2,s_1,s_2,s_3]$.
Although these degeneracies could certainly be computed on
a case-by-case basis using the orthogonality of characters,
in order to find a simple formula along the lines of eq.~(\ref{eq:centralconf})
it would first be necessary to
somehow generalize the analysis of
section 2 from SU(2) to SO(6).
We leave this intriguing problem open for future work.

\section*{Acknowledgments}

M.~S. is grateful to A.~Volovich for helpful comments.
The work of M.~S. is supported by the US National Science Foundation
under grant PHY-0638520 and by the US Department of Energy under
contract DE-FG02-91ER40688.

\end{document}